\documentclass[pra,twocolumn,10pt,superscriptaddress]{revtex4-1}
\usepackage{amssymb,amsmath,amsfonts,xcolor,graphicx}
\usepackage{times}
\usepackage{braket}
\usepackage{txfonts}
\usepackage{float}
\usepackage{lipsum}
\begin{document}

\title{ Robust nonadiabatic geometric quantum computation by dynamical correction}

\author{Ming-Jie Liang}
\affiliation{Guangdong Provincial Key Laboratory of Quantum Engineering and Quantum Materials, and School of Physics and Telecommunication\\ Engineering,  South China Normal University, Guangzhou 510006, China}

\author{Zheng-Yuan Xue} \email{zyxue83@163.com\\Published as Phys. Rev. A 106, 012603 (2022)}
\affiliation{Guangdong Provincial Key Laboratory of Quantum Engineering and Quantum Materials, and School of Physics and Telecommunication\\ Engineering,  South China Normal University, Guangzhou 510006, China}

\affiliation{Guangdong-Hong Kong Joint Laboratory of Quantum Matter and Frontier Research Institute for Physics,\\ South China Normal University, Guangzhou  510006, China}

\affiliation{Guangdong Provincial Key Laboratory of Quantum Science and Engineering, Southern University of Science and Technology,\\ Shenzhen 518055, China}

\date{\today}

\begin{abstract}
Besides the intrinsic noise  resilience property, nonadiabatic geometric phases are of  the fast evolution nature, and thus can  naturally be used in constructing quantum gates with excellent performance, i.e., the so-called nonadiabatic geometric quantum computation (NGQC). However,  previous single-loop NGQC schemes are sensitive to the operational control error, i.e., the $X$ error, due to the limitations of the implementation. Here, we propose a robust scheme for  NGQC combining with the dynamical correction technique, which still uses only simplified pulses, and thus being experimental friendly. We numerically show that our scheme can greatly improve the gate robustness in previous protocols, retaining the intrinsic merit of geometric phases. Furthermore, to fight against the dephasing noise, due to the $Z$ error,  we can incorporate the decoherence-free subspace encoding strategy. In this way,  our scheme can  be robust against both types of errors. Finally, we also propose how to implement the scheme with encoding on superconducting quantum circuits with experimentally demonstrated technology. Therefore, due to the intrinsic robustness, our scheme provides a promising alternation for the future scalable fault-tolerant quantum computation.
\end{abstract}
\maketitle

\section{Introduction}
In 1982, to simulate quantum systems, Feynman proposed the idea of building a quantum computer \cite{1982Simulating},  then Shor shows that the problem of factorizing prime numbers can be effectively solved by  using quantum computation \cite{1999Polynomial}. From then on, quantum computation is gradually being treated  as a more efficient way for dealing with some problems that are hard for classical computers. However, when manipulating quantum systems, the noise induced by their surrounding environment, as well as the operational imperfection, will inevitably lead to certain manipulation errors. Therefore, high precision   quantum manipulations are the key to realize large-scale  quantum computation  \cite{2019Nature}.

Meanwhile, together with the conventional dynamical phase, Berry observed that adiabatic and cyclic evolution of a quantum system can also lead  to  an additional  purely geometric phase, i.e., the Berry phase \cite{1984Adiabatic-Abelian}, which is Abelian, and its  non-Abelian generalization has also been proposed soon \cite{1984Adiabatic-NonAbelian}. By exploiting the global property of geometric phases, noise-resilient quantum gates can be constructed for  quantum computation \cite{1999HQC1,1999HQC2}. But, due to the required slow adiabatic evolution, considerable  gate errors will be introduced due to the decoherence. Later, Aharonov and Anandan removed the adiabatic limitation of the Berry phase, and proposed the nonabiabatic geometric phase \cite{1987NonAdiabatic-Abelian}. Then, nonadiabatic geometric quantum computation  (NGQC) protocols \cite{2001NGQC-NMR, 2002NGQC-twophaseshiftgate, 2012NHQC-ES, 2012NHQC-DFS}, constructing quantum gates using nonadiabatic geometric phases, are proposed naturally.

Recently, many different NGQC protocols have been developed \cite{,2003ZSL-twoloop-NMR,2003ZSL-singleloop,2003ZSL,2017NGQC-Ra,2018NGQC-PTC*, 2020-NGQC-Silicon*,2021-NGQC-z*}; however, after experimental verification in various quantum systems  \cite{2003NGQC-twophases-Ex,2014-SSS,2017NGQC-Continuous-variable-Ex,2020-NGQC-Sc-Ex*,2021-NGQC-Sc-Ex*,2021-NGQC-ScX-Ex}, it turns out that the advantages of geometric phase are  compromised by local noises and the limitations of the protocol.
To find better implementations, the orange-slice-shaped  scheme  \cite{2017NGQC-Ra,2018NGQC-PTC*,2020-NGQC-Silicon*,2021-NGQC-z*} is proposed with   simplified pulses, and the time and path optimal control techniques \cite{2020-NGQC-TOC-O*,2015-TOC1-O,2016-TOC2-O,2021-NGQC-SP1-O*,2021-NGQC-SP2-O*,2021-NGQC-SP3-O*}  are proposed to shorten the gate  time and thus reducing the decoherence induced error. Meanwhile, when the $Z$ error is the main error source, the dynamical decoupling method can be introduced  \cite{1999-DD,2014-NGQC-DD,2020-NGQC-DD}. However,  to enhance the gate fidelity, the requirements for the pulse parameters there are usually  strict, and thus no further pulse shaping can be allowed.  Therefore, it is necessary to find a solution that removes  pulse-shape limitations in implementing the  geometric quantum gates  while maintaining their merits.

Here, we propose a NGQC scheme with the dynamical correction technique \cite{2009NGQC-O-DECG,2012-Nature-DCG,2015-Nature-DCG,2021-NHQC-DCG*}, which has no  restriction on the pulse shape, and can greatly improve the gate robustness against the $X$ error. Meanwhile, for the $Z$ error, we can combine a decoherence-free subspace (DFS) encoding to suppress the collective dephasing noise \cite{1997-DFS1, 1998-DFS2, 1999-NoiselessCodes,2018-DFS-NGQC-XY,2019-DFS-NGQC-superadiabatic,2022-DFS-NGQC-NV}. In this way, both $X$ errors and $Z$ errors can be greatly suppressed. Finally, we propose an implementation of our proposal on superconducting quantum circuits using an experimentally demonstrated parametrically tunable coupling technique \cite{2017-PTC1,2018-PTC2,2018-PTC3,2018-PTC*}, which is numerically shown to be insensitive to both $X$ and $Z$ errors. Therefore, our scheme provides a promising strategy for realizing  robust geometric quantum  gates, which is necessary to the large-scale quantum computation.

\section{The single-loop scheme}
\begin{figure}[tbp]
	\centering
\includegraphics[width=0.95\columnwidth]{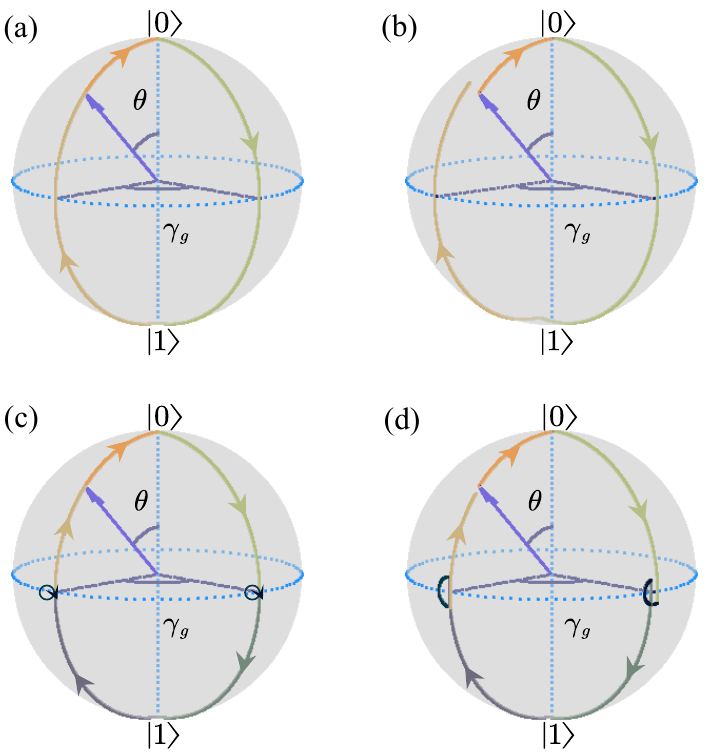}
	\caption{A schematic illustration of the evolution paths for
$H$ gate, with $\gamma_g=\pi/2$, $\theta=\pi/4$, and $\phi=0$. The evolution paths for the single-loop NGQC scheme (a) without and (b) with the systematic $X$ error. The evolution paths in our scheme (c) without and (d) with the systematic $X$ error.}\label{fig1}
\end{figure}

We first consider the  the single-loop scheme   \cite{2017NGQC-Ra, 2018NGQC-PTC*, 2020-NGQC-Silicon*, 2021-NGQC-z*}, which has no limitation on the pulse shape.
In the qubit subspace  $\{|0\rangle,|1\rangle\}$, assuming $\hbar=1$ hereafter, the general Hamiltonian for a qubit  resonantly driven by a classical driving  field is
\begin{eqnarray}
\label{Eq.5}
H_1=\frac{\Omega\left(t\right)}{2}\left(
\begin{array}{cc}
0 & e^{-i\phi} \\
e^{i\phi} & 0 \\
\end{array}
\right),
\end{eqnarray}
where $\Omega(t)$ and $\phi$ are the  amplitude and phase of the field. The geometric phase is induced by three segments that form a cyclic evolution with period $T$; the pulse areas and phases of three segments satisfy the following conditions:
\begin{eqnarray}
\int_{0}^{T_1}\Omega\left(t\right)dt &=& \theta,\quad \phi-\frac{\pi}{2}, \quad t\in\left[0,T_1\right],\nonumber\\
\int_{T_1}^{T_2}\Omega\left(t\right)dt &=& \pi,\quad \phi+\gamma_g+\frac{\pi}{2},\quad t\in\left[T_1,T_2\right],\\
\int_{T_2}^{T}\Omega\left(t\right)dt &=& \pi-\theta,\quad \phi-\frac{\pi}{2},\quad t\in\left[T_2,T\right].\nonumber
\end{eqnarray}
Note that, only the pulse areas are set here, but there is no restriction on the pulse shapes. After the cyclic evolution, one can obtain the evolution operator as
\begin{eqnarray}
\label{Eq.7}
U_S\left(T\right)&=&U_1\left(T,T_2\right)U_1\left(T_2,T_1\right)U_1\left(T_1,0\right) \nonumber\\
&=&e^{i\gamma_g\vec n\vec\sigma},
\end{eqnarray}
where $\vec n=\left(\sin{\theta}\cos{\phi},\sin{\theta}\sin{\phi},\cos{\theta}\right)$, $\vec \sigma=\left(\sigma_x,\sigma_y,\sigma_z\right)$ are the Pauli operators for the subspace $\{|0\rangle,|1\rangle\}$. Thus, by controlling the amplitude and phase of the driving field, we can obtain the appropriate parameters $\gamma_g$, $\theta$, and $\phi$ to construct any single-qubit  gate in a geometric way. The geometric illustration of the Hadamard ($H$) gate is shown in Fig. \ref{fig1}(a).

However, the above scheme is not robust against the $X$ error, in the form of $H^\epsilon\left(t\right)=\left(1+\epsilon\right)H\left(t\right)$, where $\epsilon$ is the deviation fraction of the amplitude of the driving field. Then, in the presence of the error, the evolution operator will be $U^\epsilon =e^{-i\mathcal{T}\int_{0}^{T}H^\epsilon\left(t'\right)dt'}$, where $\mathcal{T}$ is the time-ordering operator. When   $\epsilon\ll1$, the gate fidelity is defined by
\begin{eqnarray} \label{Eq.10}
F= \frac{|\textup{Tr}\left(U^\dagger U^\epsilon\right)|}{|\textup{Tr}\left(U^\dagger U\right)|}.
\end{eqnarray}
With this measure, the  fidelity of the $H$, $S$, and $T$ gates are calculated to be
\begin{eqnarray}\label{F1}
F_H&=&1-\frac{5\pi^2\epsilon^2}{32}+O\left(\epsilon^3\right), \nonumber\\
F_S&=&1-\frac{\pi^2\epsilon^2}{4}+O\left(\epsilon^3\right),\\
F_T&=&1-\frac{\left(\sqrt{2}+1\right)\pi^2\epsilon^2}{4\sqrt{2}}+O\left(\epsilon^3\right), \nonumber
\end{eqnarray}
which shows that this scheme can only suppress $X$ error up to the second order, similar to the dynamical schemes. From another point of view, it can be seen that due to the existence of the systematic $X$ error, the evolution state cannot exactly go back to the original starting point for the single-loop NGQC scheme [see Fig. \ref{fig1}(b)] and thus leads to infidelity of the constructed quantum gate.

\section{NGQC with dynamical correction}

\subsection{The scheme}

In order to solve this problem, we here propose NGQC with  dynamical correction \cite{2009NGQC-O-DECG,2012-Nature-DCG,2015-Nature-DCG,2021-NHQC-DCG*}. When the evolution state evolves along a single-loop cyclic path to the equatorial plane of the Bloch sphere, as shown in Fig.\ref{fig1}(c),
we stop the process until an additional dynamical $\pi$ pulse is finished. Note that  two additional dynamical processes will be inserted as the cyclic process arrives at the equatorial plane twice.  The Hamiltonian of the two additional inserted dynamical processes are respectively expressed as
\begin{eqnarray}
H^{I}_1\left(t\right)=\frac{1}{2}\Omega^{I}_1\left(t\right)\left(
\begin{array}{cc}
0 & e^{-i\left(\phi+\gamma\right)} \\
e^{i\left(\phi+\gamma\right)} & 0 \\
\end{array}
\right),
\end{eqnarray}
\begin{eqnarray}
H^{I}_2\left(t\right)=\frac{1}{2}\Omega^{I}_2\left(t\right)\left(
\begin{array}{cc}
0 & e^{-i\phi} \\
e^{i\phi} & 0 \\
\end{array}
\right).
\end{eqnarray}
In this way,  compared with the single-loop cyclic evolution, although the two inserted dynamical processes  bring unnecessary dynamical phase accumulation,  their summation is still zero. After the cyclic evolution period $T'$, the arbitrary single-qubit geometric  gate in Eq. (\ref{Eq.7}) can still be constructed.
After inserting two dynamical processes, the single-loop cyclic evolution with period $T'$ will be divided into seven segments; the intermediate time is defined as $T_1'$, $T_2'$, $T_3'$, $T_4'$, $T_5'$, $T_6'$, and the pulse areas and phases of the seven segments satisfy
\begin{eqnarray}
\label{Eq.15}
\int_{0}^{T_1'}\Omega\left(t\right)dt &=& \theta,\quad \phi-\frac{\pi}{2}, \quad t\in\left[0,T_1'\right],\nonumber\\
\int_{T_1'}^{T_2'}\Omega\left(t\right)dt &=& \frac{\pi}{2},\quad \phi+\gamma+\frac{\pi}{2},\quad t\in\left[T_1',T_2'\right],\nonumber\\
\int_{T_2'}^{T_3'}\Omega^I_1\left(t\right)dt &=& \pi,\quad \phi+\gamma+\pi,\quad t\in\left[T_2',T_3'\right],\nonumber\\
\int_{T_3'}^{T_4'}\Omega\left(t\right)dt &=& \frac{\pi}{2},\quad \phi+\gamma+\frac{\pi}{2}, \quad t\in\left[T_3',T_4'\right],\\
\int_{T_4'}^{T_5'}\Omega\left(t\right)dt &=& \frac{\pi}{2},\quad \phi-\frac{\pi}{2},\quad t\in\left[T_4',T_5'\right],\nonumber\\
\int_{T_5'}^{T_6'}\Omega^I_2\left(t\right)dt &=& \pi,\quad \phi,\quad t\in\left[T_5',T_6'\right],\nonumber\\
\int_{T_6'}^{T'}\Omega\left(t\right)dt &=& \frac{\pi}{2}-\theta,\quad \phi-\frac{\pi}{2},\quad t\in\left[T_6',T'\right].\nonumber
\end{eqnarray}
Note that there is still no  restriction on the pulse shapes.

\subsection{Gate performance}

Similarly, we calculate the gate fidelity under the influence of $X$ error, and the fidelities of the $H$, $S$, and $T$ gates are
\begin{eqnarray}
F'_H&=&1-\frac{\pi^2\epsilon^2}{32}+O\left(\epsilon^3\right), \nonumber\\
F'_S&=&1-\frac{\pi^4\epsilon^4}{16}+O\left(\epsilon^5\right),\\
F'_T&=&1-\frac{\left(\sqrt{2}+1\right)\pi^4\epsilon^4}{16\sqrt{2}}+O\left(\epsilon^5\right). \nonumber
\end{eqnarray}
Comparing with the single-loop scheme [see Eq. (\ref{F1})] the infidelity  of the $H$ gate  is suppressed to 1/5, and the $S$ and $T$ gate robustness   is improved from the second order to the fourth order.
We can also see from the schematic illustration of the evolution path in Fig. \ref{fig1}(d), the evolution state can still approximately  go back to the original starting point under the  influence of the $X$ error, thus leading  to robust geometric quantum gates.

\begin{figure}[tbp]
	\centering
	\includegraphics[width=\columnwidth]{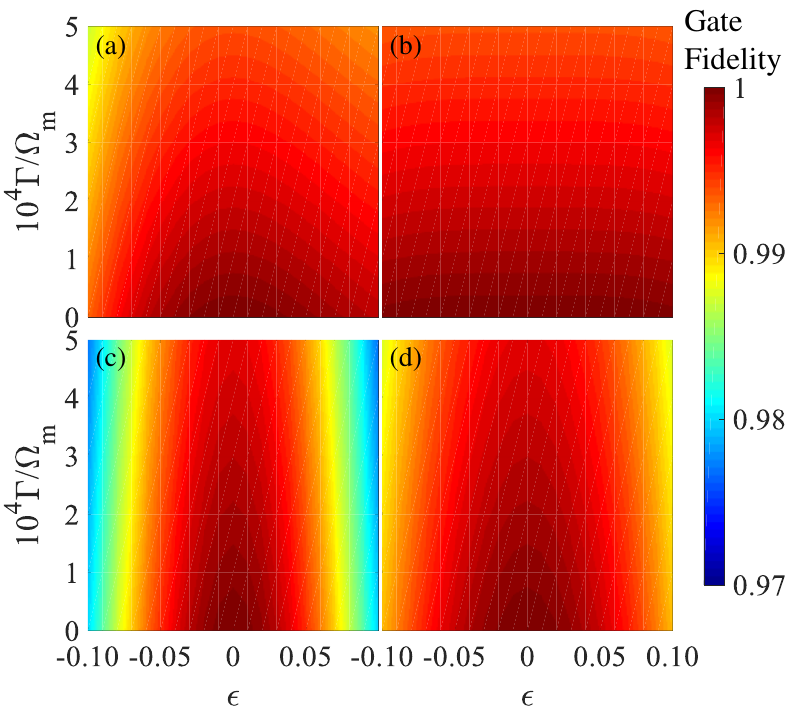}
	\caption{Gate fidelities as functions of the qubit-frequency drift fraction $\delta$ and the decoherence ratio $\Gamma/\Omega_m$. The gate robustness. (a) and (b)  are for the $H$ and $S$ gates in our scheme,   while (c) and (d) are for the $H$ and $S$ gates in the single-loop  NGQC scheme, respectively.}\label{fig2}
\end{figure}

Now, we consider a practical two-level quantum system and use the Lindblad master equation to simulate the performance of the implemented single-qubit  gate
\begin{eqnarray}
\label{mainequation}
\dot{\rho}=i\left[\rho,H\right]+\frac{1}{2}\sum_{n=1}^{2}\Gamma_n\mathcal{L}\left(\sigma_n\right),
\end{eqnarray}
where $\rho$ is the density matrix of the quantum system, and $\mathcal{L}\left(\sigma_n\right)=2\sigma_n\rho\sigma_n^\dagger-\sigma_n^\dagger\sigma_n\rho-\rho\sigma_n^\dagger\sigma_n$ is the Lindbladian of the operator $\sigma_n$, with $\sigma_1=|0\rangle\langle 1|$, $\sigma_2=|1\rangle\langle 1|-|0\rangle\langle 0|$, and $\Gamma_n=\Gamma$ is the corresponding decoherence rates. For typical quantum systems, e.g., superconducting transmon qubits \cite{2019Nature},  the ratio can usually be $\Gamma/\Omega_m \sim 10^{-4}$. Since the $H$, $S$, and $T$ gates are universal for arbitrary single-qubit gates, we take their performance for illustration purposes. They correspond a same $\phi=0$  with different $\theta$ and $\gamma$, i.e., $\theta_H=\pi/4$ and $\theta_S=\theta_T=0$;  $\gamma_H=\pi/2$, $\gamma_S=\pi/4$, and  $\gamma_T=\pi/8$, respectively. When the initial states are set to be $|\psi_i\rangle_H=|0\rangle$ and $|\psi_i\rangle_{S,T}=\left(|0\rangle+|1\rangle\right)/\sqrt{2}$, the corresponding ideal final states will be $|\psi_f\rangle_H=\left(|0\rangle+|1\rangle\right)/\sqrt{2}$, $|\psi_f\rangle_S=\left(|0\rangle+i|1\rangle\right)/\sqrt{2}$ and $|\psi_f\rangle_T=\left(|0\rangle+e^{i\pi/4}|1\rangle\right)/\sqrt{2}$.   When $\Gamma/\Omega_m=2 \times 10^{-4}$, the  state fidelities for the $H$, $S$ and $T$ gates, defined by $F^S_{H,S,T}=_{H,S,T}\langle\psi_f|\rho|\psi_f\rangle_{H,S,T}$, are  obtained as $F_H = 99.73\%$, $F_S = 99.78\%$, and $F_T = 99.79\%$, respectively.
We further compare the gate robustness  constructed by  the single-loop scheme and our scheme against $X$ error, using gate fidelity \cite{2005Fidelity}.  We take $H$ and $S$ gates for  examples,  The robustness of the $T$ gate is similar to the $S$ gate, and thus not shown here. As shown in Fig. \ref{fig2}, our scheme  shows stronger gate robustness, as predicted.

\subsection{DFS encoding}
\begin{figure}[tbp]
	\centering
	\includegraphics[width=\columnwidth]{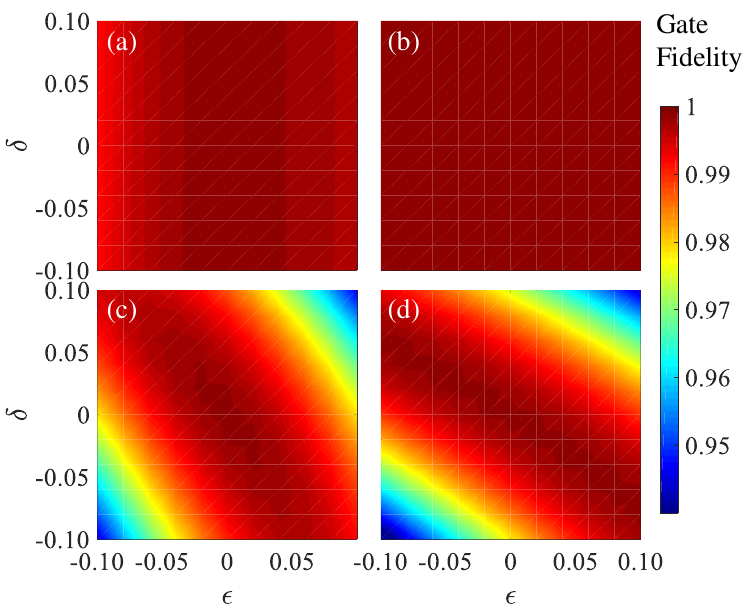}
	\caption{Gate fidelities as functions of the deviation fraction $\epsilon$ of driving amplitude and the qubit-frequency drift fraction $\delta$. The gate robustness of the $H$ and $S$ gates for our scheme with DFS encoding in  (a) and  (b), and for the single-loop   NGQC scheme in (c) and (d), respectively.}\label{fig3}
\end{figure}

Meanwhile, for the $Z$ error, we can  incorporate the DFS encoding to fight  against the collective dephasing noise. Thus, we consider two resonantly coupled two-level qubits with two-body exchange interaction as
\begin{eqnarray}
\label{L}
H_2\left(t\right)=g|1\rangle_A\langle 0|\otimes|0\rangle_B\langle 1|e^{i\phi_0}+\textup{H.c.},
\end{eqnarray}
where $g$ is coupling strengths between two physical qubits. The above Hamiltonian can be induced in many general quantum systems. Here we can take their single excitation subspace as a unit to encode the logical qubits, i.e.,
$|0\rangle_L=|10\rangle$, $|1\rangle_L=|01\rangle$,
where $|mn\rangle=|m\rangle_A\otimes|n\rangle_B$, and we use subscripts $L$ to define logical qubit states. The Hamiltonian of  Eq. (\ref{L}) in the logical basis $\{|0\rangle_L,|1\rangle_L\}$ is $H^2_L=g|0\rangle_L\langle1|e^{i\phi_0}+\textup{H.c.}$, which is in the same form as Eq. (\ref{Eq.5}).
Then, selecting the relevant parameters  according to Eq. (\ref{Eq.15}), we can obtain the geometric gate in the form of Eq. (\ref{Eq.7}).
We now consider the performance improvement of our scheme with the DFS encoding compared with single-loop NGQC. Under the influence of both the $X$ and $Z$ errors, the target Hamiltonian will be changed to $H^E\left(t\right)=\left(1+\epsilon\right)H_1\left(t\right)-\delta\Omega_m|1\rangle\langle 1|$ and $H^E_L\left(t\right)=\left(1+\epsilon\right)H^2_L\left(t\right)-\delta\Omega_m\left(|0\rangle_L\langle 0|+|1\rangle_L\langle 1|\right)$, respectively. %\textcolor[rgb]{1,0,0}{We also take $H$ and $S$ gates for  examples here.}
As shown in Fig. \ref{fig3}, the  combination of both dynamical correction and the DFS encoding lead to  greatly suppression of both $X$ and $Z$ errors.

\section{Physical implementation}
\begin{figure}[tbp]
	\centering
	\includegraphics[width=8cm]{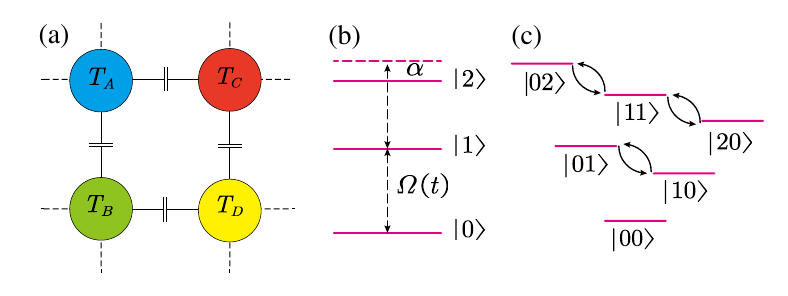}
	\caption{Illustration of our implementation scheme. (a) A 2D square qubit lattice consists of capacitively coupled superconducting transmons. (b) The energy levels of a transmon, where an  external microwave field couples the two lowest energy levels of the transmon qubit, but at the same time transitions to higher energy levels in a dispersive way. (c) The energy spectrum of two capacitively coupled transmons, different excitation subspaces can be selected by means of parametrically tunable coupling.}\label{fig4}
\end{figure}

Next, we consider the physical implementation of our scheme with DFS encoding, using capacitive coupling between adjacent transmon qubits on a two-dimensional (2D) square superconducting qubit lattice \cite{2007-2Dsquare1,2007-2Dsquare2}, as shown in Fig. \ref{fig4}(a), via parametrically tunable coupling \cite{2017-PTC1,2018-PTC2,2018-PTC3,2018-PTC*}. For two capacitively coupled transmon qubits, $T_A$ and $T_B$, the Hamiltonian can be written as
\begin{eqnarray}
H_c=\sum_{i=A,B}\sum_{j=1}^{2}\{\left[j\omega_i-\left(j-1\right)\alpha_i\right]\}\Pi_j^i \nonumber \\
+g_{AB}\prod_{k=A,B}\left(\sum_{j=1}^{2}\lambda_j\sigma_j^i\right)+\textup{H.c.},
\end{eqnarray}
where $\omega_i$ and $\alpha_i$ are the intrinsic frequency and anharmonicity of the $i$th qubit, respectively [see  Fig. \ref{fig4}(b)] and $g_{AB}$ is the fixed coupling strength between the two qubits,  $\Pi_j^i=|j\rangle_i\langle j|$, $\sigma_j^i=|j\rangle_i\langle j-1|$,  $\lambda_1=1$, and $\lambda_2=\sqrt{2}$.
To obtain an adjustable coupling strength, we add an ac driving on the  qubit $T_A$ as $\omega_A\left(t\right)=\omega_A+\varepsilon_d\sin\left(\omega_d t+\phi_d\right)$.
Converting to the interaction picture, the interaction Hamiltonian changes to
\begin{eqnarray}
\label{eq.13}
H^I_c&=&g_{AB}e^{i\beta\cos\left(\omega_d t+\phi_d\right)} \{|01\rangle\langle 10|e^{i\Delta t}\nonumber\\
& &+\sqrt{2}|11\rangle\langle 20|e^{i\left(\Delta+\alpha_A\right)t}\nonumber\\
& &+\sqrt{2}|02\rangle\langle 11|e^{i\left(\Delta-\alpha_B\right)t}\nonumber\\
& &+2|12\rangle\langle 21|e^{i\left(\Delta+\alpha_A-\alpha_B\right)t}\}+\textup{H.c.},
\end{eqnarray}
where $\Delta=\omega_A-\omega_B$, $\beta=\varepsilon_d/\omega_d$.
The single-excitation subspace $\{|01\rangle,|10\rangle\}$ can be obtained by satisfying
\begin{eqnarray}
\label{Delta_1}
\Delta=n_1\omega_d,
\end{eqnarray}
where $n_1=\pm1,\pm2,\dots$ [see Fig. \ref{fig4}(c)]. Similarly, as shown in  Fig. \ref{fig4}(c), the two-excitation subspaces $\{|11\rangle,|20\rangle\}$  and $\{|02\rangle,|11\rangle\}$ can be obtained by satisfying the conditions of
\begin{eqnarray}
\Delta+\alpha_A=n_2\omega_d,\quad \Delta-\alpha_B=n_3\omega_d,
\end{eqnarray}
respectively, where $n_2,n_3=\pm1,\pm2,\dots$.

\begin{figure}[tbp]
	\centering
	\includegraphics[width=\columnwidth]{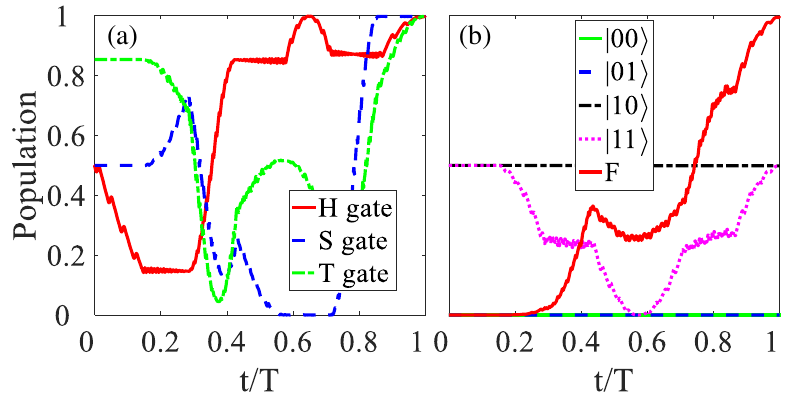}
\caption{The state dynamics of the gate fidelities of (a) single-logical
qubit $H$, $S$, and $T$ gates, and (b) two-logical qubit  controlled-Z(CZ) gate with the initial  state being $|1\rangle_L\left(|0\rangle_L+|1\rangle_L\right)/\sqrt{2}$.}
\label{fig5}
\end{figure}

For two adjacent transmon qubits, $T_A$ and $T_B$, with capacitive coupling when the driving added to $T_A$ is $\omega_A\left(t\right)=\omega_A+\varepsilon_1\sin\left(\omega_1 t+\phi_L\right)$. By meeting the condition $\Delta=\omega_1$ in Eq. (\ref{Delta_1}) with $n_1=1$, neglecting the higher-order oscillating terms,
the corresponding effective Hamiltonian is in the same as for Eq. (\ref{L})
with $g=J_1\left(\beta_1\right)g_{AB}$, $\beta_1=\varepsilon_1/\omega_1$, and $\phi_0=\pi/2-\phi_L$, and thus we can  construct arbitrary single-logical-qubit geometric gates with DFS encoding.We also use the Lindblad master equation in Eq. (\ref{mainequation}) to evaluate our implementation. Due to the limited anharmonicity for the spectrum of the transmon qubit, we set $\sigma_1'=\sum_{i=A,B}|0\rangle_i\langle1|+\sqrt{2}|1\rangle_i\langle2|$, $\sigma_2'=\sum_{i=A,B}|1\rangle_i\langle1|+2|2\rangle_i\langle2|$, i.e., we have included the qubit state leakage to the higher level.
The numerical simulations of the single-logical-qubit $H$, $S$, and $T$ gates are shown in Fig. \ref{fig5}(a), which correspond to the same $\phi_L=0$, with different $\theta_L$ and $\gamma_L$, i.e., $\theta^H_L=\pi/4$ and $\theta^S_L=\theta^T_L=0$, and   $\gamma^H_L=\pi/2$, $\gamma^S_L=\pi/4$, and  $\gamma^T_L=\pi/8$, respectively. The parameters of the physical qubits are selected as the following: the anharmonicities of the transmon qubits $T_A$ and $T_B$ are $\alpha_A=2\pi\times220$ MHz and $\alpha_B=2\pi\times245$ MHz, the parametric driving frequency equal to the frequency is difference of the two transmon qubits, i.e.,  $\Delta=\omega_1=2\pi\times700$ MHz, the decoherence rates of the two transmon qubits are  $\Gamma=2\pi\times4$ KHz,  $g_{AB}=2\pi\times20$ MHz,  and  $\beta_{1} \approx1.8$. By these settings,  the gate fidelities of the $H$, $S$, and $T$ gates can be as high as $F_L^H=99.70\%$, $F_L^S=99.77\%$, and $F_L^T=99.73\%$. Note that, combining the DFS encoding in the scheme will slightly reduce the maximal gate fidelities, but the gate robustness can be greatly improved. In particular, as we will show in the following, the the DFS encoding   scheme does not introduce additional complexity in implementing two-logical-qubit gates.

For a nontrivial two-qubit gate, we can similarly use four   physical qubits to implement nontrivial two-qubit gates. Here, we choose the transmon qubits $T_A,T_B$ and $T_C,T_D$ as the two unit elements to encode the first- and second-logical qubits; the definition of the four logical qubits and the four-dimensional subspace they encoded in DFS are
\begin{eqnarray}
S_2=\{|00\rangle_L=|1010\rangle,\quad |01\rangle_L=|1001\rangle,\nonumber\\
|10\rangle_L=|0110\rangle,\quad |11\rangle_L=|0101\rangle\},
\end{eqnarray}
where $|mnm'n'\rangle=|m\rangle_A\otimes|n\rangle_B\otimes|m'\rangle_C\otimes|n'\rangle_D$.
We choose to construct the CZ gate that generates a phase on the corresponding logical qubit $|11\rangle_L$, take one physical qubit from the two logical qubits, namely, $T_B$ and $T_D$, and couple the two physical qubits with parametrically tunable coupling, by adding the driving on the qubit $T_B$ with $\omega_B\left(t\right)=\omega_B+\epsilon_2\sin\left(\omega_2 t+\phi'_L\right)$. Similar to Eq. (\ref{eq.13}), the Hamiltonian in the interaction picture can be written as
\begin{eqnarray}
H_{BD}^I&=&g_{BD}e^{i\beta_2\cos\left(\omega_2 t+\phi'_L\right)} \{|01\rangle_{BD}\langle 10|e^{i\Delta' t}\nonumber\\
& &+\sqrt{2}|11\rangle_{BD}\langle 20|e^{i\left(\Delta'+\alpha_B\right)t}\nonumber\\
& &+\sqrt{2}|02\rangle_{BD}\langle 11|e^{i\left(\Delta'-\alpha_D\right)t}\nonumber\\
& &+2|12\rangle_{BD}\langle 21|e^{i\left(\Delta'+\alpha_B-\alpha_D\right)t}\}+\textup{H.c.},
\end{eqnarray}
where $\Delta'=\omega_B-\omega_D$, $\beta_2=\epsilon_2/\omega_2$.
By meeting the condition $\omega_2=\Delta'-\alpha_D$, we can achieve the resonant situation in the two-excitation subspace of $\{|11\rangle_{BD},|02\rangle_{BD}\}$, and after neglecting the higher-order oscillating terms by using the rotating-wave approximation, we can obtain an effective Hamiltonian of
\begin{eqnarray}
\label{H_BD_L}
H_{BD}^L=\mathcal{G}_{BD}\left(|11\rangle_L\langle A|e^{-\left(\phi'_L-{\pi}/{2}\right)}+\textup{H.c.}\right),
\end{eqnarray}
where $\mathcal{G}_{BD}=\sqrt{2}J_1\left(\beta_2\right)g_{BD}$ with $\beta_2=\epsilon_2/\omega_2$, and $|A\rangle_L=|0002\rangle$ is an auxiliary state; it does not participate in the construction of computational subspaces $S_2$. Then, we can also implement  dynamical correction nonadiabatic geometric two-qubit gates with DFS. In order to obtain the CZ gate, the corresponding gate parameters are $\phi'_L=0$, $\theta'_L=0$, and $\gamma'_L=\pi$; after the cyclic evolution period $T''$ with the  process similar to Eq. (\ref{Eq.15}),
the evolution operator in the two-logical-qubit subspace can be expressed as
\begin{eqnarray}
U_L'\left(T''\right)=
\begin{pmatrix}
1 & 0 & 0 & 0  \\
0 & 1 & 0 & 0  \\
0 & 0 & 1 & 0\\
0 & 0 & 0 & -1
\end{pmatrix},
\end{eqnarray}
in the DFS $S_2$ will be in the form of the CZ gate.
Note that, we here also only use the interaction between two physical qubits, i.e., the DFS encoding  here has not introduce additional complexity in implementing of  two-logical-qubit gates.
In order to further analyze its performance, we set the parameters of the physical qubit $T_D$ as $\alpha_D=2\pi\times245$ MHz, the coupling strength is $g_{BD}=2\pi\times20$ MHz, and the decoherence rates are $\Gamma=2\pi\times4$ KHz. For $\Delta'=\omega_{2}+\alpha_D=2\pi\times945$ MHz, $\omega_{2}=2\pi\times700$ MHz, $\beta_{2}\approx2.1$, and with an initial product state of $|1\rangle_L\left(|0\rangle_L+|1\rangle_L\right)/\sqrt{2}$, the gate fidelity of the $CZ$ gate can reach $99.56\%$, as shown in  Fig. \ref{fig5}(b).

 \section{Conclusion}
In conclusion, we propose a scheme for NGQC combined with the  dynamical correction, and   numerical simulations for the universal quantum gate set show that the dynamical correction can greatly improve the gate robustness against the $X$ error. When incorporating a DFS encoding, the influence of $Z$ error can also be suppressed. Finally, we present a physical implementation of our scheme with the DFS encoding with experimentally demonstrated techniques. Therefore, our scheme provides a promising way to achieve scalable fault-tolerant quantum computation.

\acknowledgements

This work was supported by the key-area research and Development Program of GuangDong Province (Grant No. 2018B030326001), the National Natural Science Foundation of China (Grant No. 11874156), Guangdong Provincial Key Laboratory of Quantum Science and Engineering (Grant No. 2019B121203002), Guangdong Provincial Key Laboratory (Grant no. 2020B1212060066) and
the Special Funds for the Cultivation of Guangdong College Students' Scientific and Technological Innovation ("Climbing Program" Special Funds)(Grant No. pdjh2021b0137).


\begin{thebibliography}{99}
\bibitem{1982Simulating}
R. Feynman,
Int. J. Theor. Phys.  \textbf{21}, 467  (1982).

\bibitem{1999Polynomial}
P. W. Shor,
SIAM J. Comput. \textbf{26}, 1484 (1997).

\bibitem{2019Nature}
F. Arute, K. Arya, R. Babbush, D. Bacon, J. C. Bardin, R. Barends, R. Biswas, S. Boixo, F. G.S.L. Brandao, D.A. Buell, {\it et al}.,
Nature (London) \textbf{574}, 505 (2019).

\bibitem{1984Adiabatic-Abelian}
M. V. Berry,
Proc. Roy. Soc. London A %: Mathematical, Physical and Engineering Sciences
\textbf{392}, 45 (1984).

\bibitem{1984Adiabatic-NonAbelian}
F. Wilczek and A. Zee,
Phys. Rev. Lett. \textbf{52}, 2111 (1984).

\bibitem{1999HQC1}
P. Zanardi and M. Rasetti,
Phys. Lett. A \textbf{264}, 94 (1999).

\bibitem{1999HQC2}
J. Pachos, P. Zanardi, and M. Rasetti,
Phys. Rev. A  \textbf{61}, 010305(R) (1999).

\bibitem{1987NonAdiabatic-Abelian}
Y. Aharonov and J. Anandan,
Phys. Rev. Lett. \textbf{58}, 1593 (1987).

\bibitem{2001NGQC-NMR}
Wang Xiang-Bin and M. Keiji,
Phys. Rev. Lett. \textbf{87}, 097901 (2001).

\bibitem{2002NGQC-twophaseshiftgate}
S.-L. Zhu and Z.-D. Wang,
Phys. Rev. Lett. \textbf{89}, 097902 (2002).

\bibitem{2012NHQC-ES}
E. Sj\"oqvist, D.-M. Tong, L. M. Andersson, L. M. Andersson, B. Hessmo, M. Johansson, and K. Singh,
New J. Phys. \textbf{14} 103035 (2012),

\bibitem{2012NHQC-DFS}
G.-F. Xu, J. Zhang, D.-M. Tong, E. Sj\"oqvist, and L. C. Kwek,
Phys. Rev. Lett. \textbf{109}, 170501 (2012).

\bibitem{2003ZSL-twoloop-NMR}
S.-L. Zhu and Z.-D. Wang,
Phys. Rev. A \textbf{67}, 022319 (2003).

\bibitem{2003ZSL-singleloop}
X.-D. Zhang, S.-L. Zhu, L. Hu, and Z.-D. Wang,
Phys. Rev. A \textbf{71}, 014302 (2005).

\bibitem{2003ZSL}
S.-L. Zhu and Z.-D. Wang,
Phys. Rev. Lett. \textbf{91}, 187902 (2003).

\bibitem{2017NGQC-Ra}
P.-Z. Zhao, X.-D. Cui, G.-F. Xu, E. Sj\"oqvist, and D.-M. Tong,
Phys. Rev. A \textbf{96}, 052316 (2017).

\bibitem{2018NGQC-PTC*}
T. Chen and Z.-Y. Xue,
Phys. Rev. Applied \textbf{10}, 054051 (2018).

\bibitem{2020-NGQC-Silicon*}
C. Zhang, T. Chen, S. Li, X. Wang, and Z.-Y. Xue,
Phys. Rev. A \textbf{101}, 052302 (2020).

\bibitem{2021-NGQC-z*}
J. Zhou, S. Li, G.-Z. Pan, G. Zhang, T. Chen, and Z.-Y. Xue,
Phys. Rev. A \textbf{103} 032609 (2021).

\bibitem{2003NGQC-twophases-Ex}
D. Leibfried, B. Demarco, V. Meyer, D. Lucas, M. Barrett, B. J. I. Wm, B. Jelenkovic, C. Langer, T. Rosenband, and D. J. Wineland,
Nature (London) \textbf{422}, 412 (2003).

\bibitem{2014-SSS}
C. Zu, W.-B. Wang, L. He, W.-G. Zhang, C.-Y. Dai, F. Wang, and L.-M. Duan,
Nature  (London)  \textbf{514}, 72 (2014).

\bibitem{2017NGQC-Continuous-variable-Ex}
C. Song, S.-B. Zheng, P. Zhang, K. Xu, L. Zhang, Q. Guo, W. Liu, D. Xu, H. Deng, and K. Huang,
Nat. Commun. \textbf{8}, 1061 (2017).

\bibitem{2020-NGQC-Sc-Ex*}
Y. Xu, Z. Hua, T. Chen, X. Pan, X. Li, J. Han, W. Cai, Y. Ma, H. Wang, Y. P. Song, Z.-Y. Xue, and L. Sun, Phys. Rev. Lett. \textbf{124}, 230503 (2020).

\bibitem{2021-NGQC-Sc-Ex*}
S. Li, B.-J. Liu, Z. Ni, L. Zhang, Z.-Y. Xue, J. Li, F. Yan, Y. Chen, S. Liu, M.-H. Yung, Y. Xu, and D. Yu,
Phys. Rev. Applied \textbf{16}, 064003 (2021).

\bibitem{2021-NGQC-ScX-Ex}
P.-Z. Zhao, Z.-J. Dong, Z.-X. Zhang, G.-P. Guo, and Y. Yin,
Sci. China Phys. Mech. Astron. \textbf{64},  {250362} (2021).

\bibitem{2020-NGQC-TOC-O*}
T. Chen and Z.-Y. Xue,
Phys. Rev. Applied \textbf{14}, 064009 (2020).

\bibitem{2015-TOC1-O}
X.-T. Wang, M. Allegra, K. Jacobs, S. Lloyd, C. Lupo, and M. Mohseni,
Phys. Rev. Lett. \textbf{114}, 170501 (2015).

\bibitem{2016-TOC2-O}
J.-P. Geng, Y. Wu, X.-T. Wang, K.-B. Xu, F.-Z. Shi, Y.-J. Xie, X. Rong, and J.-F. Du, Phys. Rev. Lett. \textbf{117}, 170501 (2016).

\bibitem{2021-NGQC-SP1-O*}
S. Li, J. Xue, T. Chen, and Z.-Y. Xue,
Adv. Quantum Technol. \textbf{4}, 2000140 (2021).

\bibitem{2021-NGQC-SP2-O*}
C.-Y. Ding, L.-N. Ji, T. Chen, and Z.-Y. Xue,
Quantum Sci. Technol. \textbf{7}, 015012 (2021).

\bibitem{2021-NGQC-SP3-O*}
C.-Y. Ding, Y. Liang, K.-Z. Yu, and Z.-Y. Xue,
Appl. Phys. Lett. \textbf{119}, 184001 (2021).

\bibitem{1999-DD}
L. Viola, E. Knill, and S. Lloyd,
Phys. Rev. Lett. \textbf{82}, 2417 (1999).

\bibitem{2014-NGQC-DD}
G.-F. Xu and G.-L. Long,
Phys. Rev. A \textbf{90}, 022323 (2014).

\bibitem{2020-NGQC-DD}
X. Wu and P.-Z. Zhao,
Phys. Rev. A \textbf{102}, 032627 (2020).

\bibitem{2009NGQC-O-DECG}
K. Khodjasteh and L. Viola,
Phys. Rev. Lett. \textbf{102}, 080501 (2009).

\bibitem{2012-Nature-DCG}
X. Wang, L. S. Bishop, J. P. Kestner, E. B. Barnes, K. Sun, and S. D. Sarma,
Nat. Commun. \textbf{3}, 997 (2012).

\bibitem{2015-Nature-DCG}
X. Rong, J.-P. Geng, F.-Z. Shi, Y. Liu, K.-B. Xu, W.-C. Ma, F. Kong, Z. Jiang, Y. Wu, and J.-F. Du,
Nat. Commun. \textbf{6}, (2015).

\bibitem{2021-NHQC-DCG*}
S. Li and Z.-Y. Xue,
Phys. Rev. Applied \textbf{16}, 044005 (2021).

\bibitem{1997-DFS1}
L.-M. Duan and G.-C. Guo,
Phys. Rev. Lett. \textbf{79}, 1953 (1997).

\bibitem{1998-DFS2}
D. A. Lidar, I. L. Chuang, and K. B. Whaley,
Phys. Rev. Lett. \textbf{81}, 2594 (1998).

\bibitem{1999-NoiselessCodes}
P. Zanardi and M. Rasetti,
Phys. Rev. Lett. \textbf{79}, 3306 (1997).


\bibitem{2018-DFS-NGQC-XY}
V. A. Mousolou,
Europhys. Lett. \textbf{121}, 20004 (2018).

\bibitem{2019-DFS-NGQC-superadiabatic}
J.-Z. Li, Y.-X. Du, Q.-X. Lv, Z.-T. Liang, W. Huang and H. Yan,
Quantum Inf. Process. \textbf{18}, 17 (2019).

\bibitem{2022-DFS-NGQC-NV}
M.-R. Yun, F.-Q. Guo, L.-L. Yan, E.-J. Liang, Y. Zhang, S.-L. Su, C.-X. Shan and Y. Jia,
Phys. Rev. A \textbf{105}, 012611 (2022).

\bibitem{2017-PTC1}
M. Roth, M. Ganzhorn, N. Moll, S. Filipp, G. Salis, and S. Schmidt, Phys. Rev. A \textbf{96}, 062323 (2017).

\bibitem{2018-PTC2}
M. Reagor, C. B. Osborn, N. Tezak, A. Staley, G. Prawiroatmodjo, M. Scheer, N. Alidoust, E. A. Sete, N. Didier, M. P. da Silva, {\it et al}.,
Sci. Adv. \textbf{4}, eaao3603 (2018).

\bibitem{2018-PTC3}
S. A. Caldwell, N. Didier, C. A. Ryan, E. A. Sete, A. Hudson, P. Karalekas, R. Manenti, M. P. da Silva, R. Sinclair, E. Acala, {\it et al}.,
Phys. Rev. Appl. \textbf{10}, 034050 (2018).

\bibitem{2018-PTC*}
X. Li, Y. Ma, J. Han, T. Chen, Y. Xu, W. Cai, H. Wang, Y.-P. Song, Z.-Y. Xue, Z.-Q. Yin, and L. Sun,
Phys. Rev. Applied {\bf 10}, 054009 (2018).

\bibitem{2005Fidelity}
A. Klappenecker and M. Roetteler,
{Proc. IEEE Int. Symp. Inf. Theory} \textbf{76}, 1740 (2005).

\bibitem{2007-2Dsquare1}
J. Koch, T. M. Yu, J. Gambetta, A. A. Houck, D. I. Schuster, J. Majer, A. Blais, M. H. Devoret, S. M. Girvin, and R. J. Schoelkopf, Phys. Rev. A \textbf{2005}, 042319 (2007).

\bibitem{2007-2Dsquare2}
J. Q. You, X. Hu, S. Ashhab, and F. Nori,
Phys. Rev. B \textbf{75}, 140515(R) (2007).

\end{thebibliography}
\end{document}